# Mining Fuzzy β-Certain and β-Possible rules from incomplete quantitative data by rough sets

A.S,Mohammadi
Department of Information Technology, Payame Noor University,
PO BOX 19395-3697 Tehran, IRAN
Ali_soultanmohammadi@yahoo.com

L. Asadzadeh
Department of Information Technology, Payame Noor University,
PO BOX 19395-3697 Tehran, IRAN
l_asadzadeh@pnu.ac.ir

D.D,Rezaee
Department of Information Technology, Payame Noor University,
PO BOX 19395-3697 Tehran, IRAN
D_dashchi_rezaee@yahoo.com

*Abstract*─The rough-set theory proposed by Pawlak, has been widely used in dealing with data classification problems. The original rough-set model is, however, quite sensitive to noisy data. Tzung thus proposed deals with the problem of producing a set of fuzzy certain and fuzzy possible rules from quantitative data with a predefined tolerance degree of uncertainty and misclassification. This model allowed , which combines the variable precision rough-set model and the fuzzy set theory, is thus proposed to solve this problem. This paper thus deals with the problem of producing a set of fuzzy certain and fuzzy possible rules from incomplete quantitative data with a predefined tolerance degree of uncertainty and misclassification. A new method, incomplete quantitative data for rough-set model and the fuzzy set theory, is thus proposed to solve this problem. It first transforms each quantitative value into a fuzzy set of linguistic terms using membership functions and then finding incomplete quantitative data with lower and the fuzzy upper approximations. It second calculates the fuzzy β-lower and the fuzzy β-upper approximations. The certain and possible rules are then generated based on these fuzzy approximations. These rules can then be used to classify unknown objects.

## I. Introduction

Machine learning and data mining techniques have recently been developed to find implicitly meaningful pat-terns and ease the knowledge-acquisition bottleneck. Among these approaches, deriving inference or association rules from training examples is the most common [9,12,13]. Given a set of examples and counterexamples of a concept, the learning program tries to induce general rules that describe all or most of the positive training instances and none or few of the counterexamples [6,7]. If the training instances belong to more than two classes, the learning program tries to induce general rules that describe each class. Recently, the rough-set theory has been used in reasoning and knowledge acquisition for expert systems [4,14]. It was proposed by Pawlak in 1982 [19], with the concept of equivalence classes as its basic principle. Several applications and extensions of the rough-set theory have also been proposed. Examples are [14,18] reasoning with incomplete information,[2] knowledge-base reduction, [10] data mining, [16] rule discovery. Due to the success of the rough-set theory to knowledge acquisition, many researchers in database and machine learning fields are interested in this new research topic because it offers opportunities to discover useful information in training examples. [6,17] mentioned that the main issue in the rough-set approach was the formation of good rules.

## II. Review of the variable precision rough-set model

The rough-set theory, proposed by Pawlak in 1982, can serve as a new mathematical tool to deal with data classification problems [19]. It adopts the concept of equivalence classes to partition the training instances according to some criteria. Two kinds of partitions are formed in the mining process: lower approximations and upper approximations, from which certain and possible rules are easily derived. Let X be an arbitrary subset of the universe U, and B be an arbitrary subset of the attribute set A. The lower and the upper approximations for B on X, denoted $B_*(X)$ and $B^*(X)$ respectively, are defined as follows:

$$B_*(X) = \{x | x \in U, B(X) \subseteq X\}, \qquad (1)$$

$$B^*(X) = \{x | x \in U \text{ and } B(X) \cap X \neq \emptyset\}. \qquad (2)$$

Elements in $B_*(x)$ can be classified using attribute set B as members of the set X with full certainty, and $B_*(x)$ is thus called a lower approximation of X. On the other hand, elements in B(x) can be classified using attribute set B as members of the set X only with partial certainty, and $B^*(x)$ is thus called an upper approximation of X. The original rough-set model is quite sensitive to noisy data [7]. When noisy data exists, the lower and the upper approximations cannot normally be formed [17]. Let two sets X and Y be non-empty subsets of the universal set (X, Y,

U). The rough inclusion function is then defined as follows:

$$\mu(X, Y) = \frac{\text{card}(X \cap Y)}{card(X)} \quad (3)$$

If the rough inclusion value equals to 1, then X is totally included in Y (X,Y). Otherwise, the rough inclusion value ranges between 0 to 1, and X is said partially included in Y. Also, the relative degree of misclassification of the set X with respect to set Y is defined as :

$$c(X, Y) = 1 - \frac{\text{card}(X \cap Y)}{card(X)} \quad (4)$$

Based on the relative degree of misclassification, in the[17], generalized the lower and upper approximations of the original rough-set model with a majority inclusion threshold β. The β-lower and the β-upper approximations are defined as follows:

$$B_{*\beta}(X) = \{x | x \in U, c(B(x),X) \leq \beta \} \quad (5)$$
$$B^{*}_{\beta}(X) = \{x | x \in U, c(B(x),X) < 1- \beta \}. \quad (6)$$

### III. Incomplete data sets

Data sets can be roughly classified into two classes: complete and incomplete data sets. All the objects in a complete data set have known attribute values. If at least one object in a data set has a missing value, the data set is incomplete. the symbol '*' denotes an unknown attribute value. Learning rules from incomplete data sets is usually more difficult than from complete data sets. In the past, several methods were proposed to handle the problem of incomplete data sets [21]. For example, incomplete data sets may be transformed into complete data sets by similarity measures or by removing objects with unknown values before learning programs begin. Incomplete data sets may also be directly processed in a particular way to get the rules [20,22]. In the [20] proposed a rough-set approach to directly learn rules from incomplete data sets without guessing unknown attribute values. In the [22] modified [20] approach by introducing the rough entropy to distinguish the power of the attribute subsets that have the same partition for similarity relations.

### IV. Notation

Notation used in this paper is described as follows:
- U    universe of all objects
- β    tolerance degree of noise and misclassification
- n    total number of training examples (objects) in U
- $Obj^{(i)}$  ith training example (object), $1 \leq i \leq n$
- A    set of all attributes describing U
- m    total number of attributes in A
- B    an arbitrary subset of A
- $A_j$    j th attribute, $1 \leq j \leq m$
- $|A_j|$  number of fuzzy regions for $A_j$
- $R_{jk}$  k th fuzzy region of $A_j$, $1 \leq k \leq |A_j|$
- $v_j^{(i)}$  quantitative value of $A_j$ for $Obj^{(i)}$
- $f_j^{(i)}$  fuzzy set converted from v
- $f_{jk}^{(i)}$  membership value of v in region R
- C    set of classes to be determined
- c    total number of classes in C
- $X_l$  l th class, $1 \leq l \leq c$
- $B(Obj^{(i)})$  the fuzzy incomplete equivalence classes in which $Obj^{(i)}$ exists
- $B_k^c(Obj^{(i)})$  the certain part of the kth fuzzy incomplete equivalence class in $B(Obj^{(i)})$
- $B_*(X)$  the fuzzy incomplete lower approximation for B on X
- $B^*(X)$  the fuzzy incomplete upper approximation for B on X

When the same linguistic term $R_{jk}$ of an attribute $A_j$ exists in two fuzzy objects $Obj^{(i)}$ and $Obj^{(r)}$ with membership values $f_{jk}^{(i)}$ and $f_{jk}^{(r)}$ larger than zero, $Obj^{(i)}$ and $Obj^{(r)}$ are said to have a fuzzy indiscernibility relation (or fuzzy equivalence relation) on attribute $A_j$ with a membership value equal to min ($f_{jk}^{(i)} \cap f_{jk}^{(r)}$). Also, if the same linguistic terms of an attribute subset B exist in both $Obj^{(i)}$ and $Obj^{(r)}$ with membership values larger than zero, $Obj^{(i)}$ and $Obj^{(r)}$ are said to have a fuzzy indiscernibility relation (or a fuzzy equivalence relation) on attribute subset B with a membership value equal to the minimum of all the membership values .These fuzzy equivalence relations thus partition the fuzzy object set U into several fuzzy subsets that may overlap, and the result is denoted by U/B. The set of fuzzy partitions, based on B and including $Obj^{(i)}$, is denoted $B(Obj^{(i)})$. Thus, $B(Obj^{(i)}) = \{((B_1(Obj^{(i)}), \mu_{B1}(Obj^{(i)})) \ldots ((B_r(Obj^{(r)}), \mu_{Br}(Obj^{(r)}))$, where r is the number of partitions included in $B(Obj^{(i)})$, $B_j(Obj^{(i)})$ is the jth partition in $B(Obj^{(i)})$, and $\mu_{Bj}(Obj^{(r)})$ is the membership value of the jth partition.

Since an incomplete quantitative data set contains unknown attribute values, each object $Obj^{(i)}$ is thus represented as a tuple ($Obj^{(i)}$, symbol), where the symbol may be certain (c) or uncertain (u). If an object $Obj^{(i)}$ has an uncertain value for attribute $A_j$, then ($Obj^{(i)}$, u) is put in each fuzzy equivalence class of attribute $A_j$. The fuzzy sets formed in this way are called fuzzy incomplete equivalence classes, which are not necessarily equivalence classes. The fuzzy incomplete lower and upper approximations for B on X, denoted $B_*(X)$ and $B^*(X)$ respectively, are defined as follows:

$$B_*(X_l) = \{(B_k(Obj^{(i)}), \mu_{Bk}(Obj^{(i)})) | 1 \leq i \leq n, obj^{(i)} \in X_l, B_k^c(Obj^{(i)}) \subseteq X_l, 1 \leq k \leq |B(Obj^{(i)})|\}. \quad (7)$$

$B^*(X_l) = \{(B_k(Obj^{(i)}), \mu_{Bk}(Obj^{(i)})) | 1 \leq i \leq n, B_k{}^c(Obj^{(i)}) \cap Xl \neq \emptyset, B_k{}^c(Obj^{(i)}) \not\subset X_l$
$1 \leq k \leq |B(Obj^{(i)})|\}.$ (8)

Here, the definition of the fuzzy incomplete upper approximation has the constraint $B_k{}^c(Obj^{(i)}) \not\subset X$ to exclude the objects in the fuzzy incomplete lower approximation for avoiding redundant calculation. The fuzzy β-lower and the fuzzy β-upper approximations for B on X, denoted $B_{*\beta}(X)$ and $B^*{}_\beta(X)$ respectively, are defined as follows:

$B_{*\beta}(X_l) = \{(B_k(x), \mu_{Bk}(x)) | x \in U, c(B_k(x), X) \leq \beta, 1 \leq k \leq |B(x)|\}$ (9)

$B^*{}_\beta(X_l) = \{(B_k(x), \mu_{Bk}(x)) | x \in U, \beta < c(B_k(x), X) \leq 1-\beta, 1 \leq k \leq |B(x)|\}.$ (10)

After the fuzzy β-lower and the fuzzy β-upper approximations are found, both β-certain and β-uncertain rules can thus be derived.

## V. The proposed algorithm for incomplete quantitative data sets

In the section, a learning algorithm based on rough sets is proposed, which can simultaneously estimate the missing values and derive fuzzy certain and possible rules from incomplete quantitative data sets. The proposed fuzzy learning algorithm first transforms each quantitative value into a fuzzy set of linguistic terms using membership functions. The details of the proposed fuzzy learning algorithm are described as follows.

The fuzzy mining algorithm for β-certain and β-possible rules:

Input: A incomplete quantitative data set with n objects, each with m attribute values, β is tolerance degree of noise and misclassification, and a set of membership functions.

Output: A set of maximally general fuzzy β-certain and fuzzy β-possible rules.

Step 1: Partition the object set into disjoint subsets according to class labels. Denote each set of objects belonging to the same class $C_l$ as $X_l$.

Step 2: Transform the quantitative value $v_j^{(i)}$ of each object $Obj^{(i)}$; i =1 to n, for each attribute $A_j$; j = 1 to m, into a fuzzy set $f_j^{(i)}$, represented as,

$$\frac{f_{j1}^{(i)}}{R_{j1}} + \frac{f_{j2}^{(i)}}{R_{j2}} + ... + \frac{f_{jl}^{(i)}}{R_{jl}}$$ (11)

using the given membership functions, where $R_{jk}$ is the kth fuzzy region of attribute $A_j$; $f_{jk}^{(i)}$ is $v_j^{(i)}$'s fuzzy membership value in region $R_{jk}$, and l (= $|A_j|$) is the number of fuzzy regions for $A_j$. If $Obj^{(i)}$ has a missing value for $A_j$, keep it with a missing value (*).

Step 3: Find the fuzzy incomplete elementary sets of singleton attributes; That is, if an object $Obj^{(i)}$ has a certain fuzzy membership value $f_{jk}^{(i)}$ for attribute $A_j$, put ($Obj^{(i)}$, c) into the fuzzy incomplete equivalence class from $A_j = R_{jk}$; If $Obj^{(i)}$ has a missing value for $A_j$, put ($Obj^{(i)}$, u) into each fuzzy incomplete equivalence class from $A_j$; The membership value $\mu_{Ajk}$ of a fuzzy incomplete class for $A_j = R_{jk}$ is calculated as:

$$\mu_{A_{jk}} = \text{Min } f_{jk}^{(i)}$$ (12)

where $Obj^{(i)}$ is certain and $f_{jk}^{(i)} \neq 0$.

Step 4: Initialize q = 1, where q is used to count the number of attributes currently being processed for fuzzy incomplete lower approximations.

Step 5: Compute the fuzzy incomplete lower approximations of each subset B with q attributes for each class $X_l$ as:

$B_*(X_l) = \{(B_k(Obj^{(i)}), \mu_{Bk}(Obj^{(i)})) | 1 \leq i \leq n, obj^{(i)} \in X_l, B_k{}^c(Obj^{(i)}) \subseteq Xl, 1 \leq k \leq |B(Obj^{(i)})|\}.$ (13)

where $B(Obj^{(i)})$ is the set of fuzzy incomplete equivalence classes including $Obj^{(i)}$ and derived from attribute subset B, $B_k{}^c(Obj^{(i)})$ is the certain part of the kth fuzzy incomplete equivalence class in $B(Obj^{(i)})$.

Step 6: If $Obj^{(i)}$ exists in fuzzy incomplete equivalence class $B_k{}^c(Obj^{(i)})$ of the kth region combination $R_B^k$ from attribute subset B in a fuzzy incomplete lower approximation, assign the uncertain value of $Obj^{(i)}$ as:

$$\frac{\sum_{obj^{(r)} \in B_k^c(obj^{(i)})} v_j^{(r)} \cdot f_{jk}^{(r)}}{\sum_{obj^{(r)} \in B_k^c(obj^{(i)})} f_{jk}^{(r)}}$$ (14)

where $v_j^{(r)}$ is the quantitative value of $Obj^{(r)}$ for attribute $A_j$ and $f_{jk}^{(r)}$ is $v_j^{(r)}$'s fuzzy membership value in $R_B^k$. Also transform the estimated $Obj^{(i)}$ value into a fuzzy set, remove ($Obj^{(i)}$, u)'s with membership values equal to zero from the fuzzy incomplete equivalence classes, change ($Obj^{(i)}$, u)'s with membership values not equal to zero into ($Obj^{(i)}$, c)'s and re-calculate the membership values of the fuzzy incomplete equivalence classes including them by the minimum operation. Besides, backtrack to the previously found fuzzy incomplete lower approximations for doing the same actions on $Obj^{(i)}$.
Step 7: Set q = q+1 and repeat Steps 5–7 until q > m.

Step 8: Initialize h = 1, where h is used to count the number of attributes currently being processed for fuzzy incomplete upper approximations.

Step 9: Compute the fuzzy incomplete upper approximations of each subset B with q attributes for each class $X_l$

$$B^*(X_l) = \{(B_k(Obj^{(i)}), \mu_{Bk}(Obj^{(i)}))| \leq l \leq n, B_k{}^c(Obj^{(i)}) \cap X_l \neq \emptyset, B_k{}^c(Obj^{(i)}) \not\subset X_l, 1 \leq k \leq |B(Obj^{(i)})|\}. \quad (15)$$

where $B(Obj^{(i)})$ is the set of fuzzy incomplete equivalence classes including $Obj^{(i)}$ and derived from attribute subset B, $B_k{}^c(Obj^{(i)})$ is the certain part of the kth fuzzy incomplete equivalence class in $B(Obj^{(i)})$.

Step 10: Do the following sub steps for each uncertain instance $Obj^{(i)}$ in the fuzzy incomplete upper approximations:

a) If $Obj^{(i)}$ exists in fuzzy incomplete equivalence class $B_k{}^c(Obj^{(i)})$ of the kth region combination $R_B{}^k$ from attribute subset B in a fuzzy incomplete lower approximation, assign the uncertain value of $Obj^{(i)}$ as:

$$\frac{\sum_{obj^{(r)} \in B_k^c(obj^{(i)}) \& obj^{(r)} \in X_l} v_j^{(r)} \cdot f_{jk}^{(r)}}{\sum_{obj^{(r)} \in B_k^c(obj^{(i)}) \& obj^{(r)} \in X_l} f_{jk}^{(r)}} \quad (16)$$

where $v_j^{(r)}$ is the quantitative value of $Obj^{(r)}$ for attribute $A_j$ and $f_{jk}^{(r)}$ is $v_j^{(r)}$'s fuzzy membership value in $R_B{}^k$. Also transform the estimated $Obj^{(i)}$ value into a fuzzy set, remove $(Obj^{(i)}, u)$'s with membership values equal to zero from the fuzzy incomplete equivalence classes, change $(Obj^{(i)}, u)$'s with membership values not equal to zero into $(Obj^{(i)}, c)$'s and re-calculate the membership values of the fuzzy incomplete equivalence classes including them by the minimum operation. Besides, backtrack to the previously found fuzzy incomplete lower approximations for doing the same actions on $Obj^{(i)}$.

b) If an object $Obj^{(i)}$ still exists in more than one fuzzy incomplete equivalence class in a fuzzy incomplete upper approximation, use the equivalence class with the maximum plausibility measure to estimate the uncertain value of $Obj^{(i)}$. The estimation and processing are the same as those stated in Step 10(a). Calculate the plausibility measures of each fuzzy incomplete equivalence class in an upper approximation for each class $X_l$ as:

$$P(B_k^c(obj^{(i)})) = \frac{\sum_{obj^{(r)} \in B_k^c(obj^{(i)}) \& obj^{(r)} \in X_l} \mu_{jk}^{(r)}}{\sum_{obj^{(r)} \subset B_k^c(obj^{(i)})} \mu_{jk}^{(r)}} \quad (17)$$

c) If an object $Obj^{(i)}$ exists in more than one fuzzy incomplete equivalence class in a fuzzy incomplete upper approximation and them plausibility measure together are equal, define for determine uncertain value of $Obj^{(i)}$ in it equivalence class that number certain objects that more of classes other. If number certain objects classes are equal, hence define that class than are include least label.

Step 11: Set h=h+1 and repeat Steps 9–11 until h > m.

Step 12 : Initialize l= 1, where l is used to count the number of classes being processed.

Step 13: Calculate the relative degree of misclassification of each attribute subset $B_k$ for class $X_l$ as:

$$c(B_k(x), X_l) = 1 - \frac{\sum_{y \in (B_k(x) \cap X_l)} \mu_{B_k}(y)}{\sum_{y \in B_k(x)} \mu_{B_k}(y)} \quad (18)$$

Step 14: Calculate the modified fuzzy β-lower and β-upper approximation of each attribute subset B for class $X_l$ as equation "(9)" and "(10)".

Step 15: Derive the β-certain rules from the fuzzy β-lower approximation $B_{*\beta}(X_l)$ of any subset B, set the membership values in the β-lower approximation as the effectiveness for future data, and calculate the plausibility measure of each rule for $B_k(x)$ as :

$$1 - c(B_k(x), X_l) \quad (19)$$

Step 16: Remove the β-certain rules with the condition parts more specific and effectiveness measure equal to or smaller than those of some other β-certain rules.

Step 17: Derive the β-possible rules from the fuzzy β-upper approximation $B^*{}_\beta(X)$ of any subset B, set the membership values in the β-upper approximation as the effectiveness for future data, and calculate the plausibility measure of each rule for $B_k(x)$ as equation "(19)".

Step 18: Remove the β-possible rules with the condition parts more specific and both the effectiveness and plausibility equal to or smaller than those of some other β-certain or β-possible rules.

Step 19: Set l=l+1 and Repeat Steps 13 to 19 until l≤ c.

Step 20 : Output the β-certain and β-possible fuzzy rules .

## VI. an example

In this section, an example is given to show how the proposed algorithm can be used to generate maximally general fuzzy β-certain and fuzzy β-possible rules from incomplete quantitative data. Table 1 shows an incomplete quantitative data set. Assume the membership functions for each attribute are given by experts as shown in Fig. 1. The proposed learning algorithm processes this incomplete quantitative data set as follows.

Table 1: The quantitative data set as an Example

| Object | Systolic pressure (SP) | Diastolic pressure (DP) | Blood pressure (BP) |
|---|---|---|---|
| $Obj^{(1)}$ | 122 | 80 | N |
| $Obj^{(2)}$ | 155 | 90 | H |
| $Obj^{(3)}$ | 130 | 92 | N |
| $Obj^{(4)}$ | 87 | 68 | L |
| $Obj^{(5)}$ | * | 93 | H |
| $Obj^{(6)}$ | 150 | 100 | H |
| $Obj^{(7)}$ | 95 | * | L |

Step 1: Since three classes exist in the data set, three partitions are thus formed as follows:

$X_L = \{obj^{(4)}, obj^{(7)}\}$,
$X_N = \{obj^{(1)}, obj^{(3)}\}$,
$X_H = \{obj^{(2)}, obj^{(5)}, obj^{(6)}\}$.

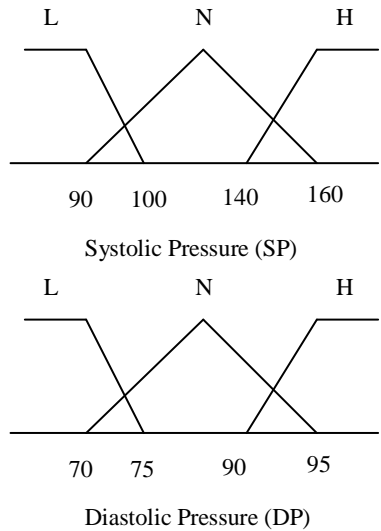

Fig. 1. The given membership functions of each attribute.

Step 2: The quantitative values of each object are transformed into fuzzy sets. Take the attribute Systolic Pressure (SP) in $Obj^{(2)}$ as an example. The value ''155'' is converted into a fuzzy set (0.1/N + 0.75/H) using the given membership functions. Results for all the objects are shown in Table 2.

Table 2: The fuzzy set transformed from the data in Table 1

| Object | Systolic pressure (SP) | Diastolic pressure (DP) | Blood pressure (BP) |
|---|---|---|---|
| $Obj^{(1)}$ | 0.9/N | 0.9/N | N |
| $Obj^{(2)}$ | 0.1/N+0.75/H | 0.4/N | H |
| $Obj^{(3)}$ | 0.85/N | 0.3/N+0.4/H | N |
| $Obj^{(4)}$ | 1/L | 1/L | L |
| $Obj^{(5)}$ | * | 0.16/N+0.6/H | H |
| $Obj^{(6)}$ | 0.3/N+0.5/H | 1/H | H |
| $Obj^{(7)}$ | 0.5/L+0.1/N | * | L |

Step 3: The elementary fuzzy sets of the singleton attributes SP and DP are found as follows:

U/{SP} = {{(obj$^{(1)}$,c)(obj$^{(2)}$,c)(obj$^{(3)}$,c)
(obj$^{(6)}$,c)(obj$^{(7)}$,c)(obj$^{(5)}$,u),0.1}
{(obj$^{(2)}$,c)(obj$^{(6)}$,c)(obj$^{(5)}$,u),0.5}
{(obj$^{(4)}$,c)(obj$^{(7)}$,c)(obj$^{(5)}$,u),0.5}}

and

U/{DP} = {{(obj$^{(1)}$,c)(obj$^{(2)}$,c)
(obj$^{(3)}$,c)(obj$^{(7)}$,u)(obj$^{(5)}$,c),0.16}
{(obj$^{(3)}$,c)(obj$^{(5)}$,c)(obj$^{(6)}$,c)(obj$^{(7)}$,u),0.4}
{(obj$^{(4)}$,c)(obj$^{(7)}$,u),0.5}}

Step 4: q is initially set at 1, where q is used to count the number of the attributes currently being processed for fuzzy incomplete lower approximations.

Step 5: The fuzzy incomplete lower approximation of one attribute for $X_H$={$Obj^{(2)}$, $Obj^{(5)}$, $Obj^{(6)}$} is first calculated. Since only the certain part ($Obj^{(2)}$,c) of the first incomplete equivalence class for attribute SP is included in $X_H$ and the uncertain instances $Obj^{(5)}$ belong to $X_H$, thus:

$SP_*(X_H) = \{(obj^{(2)},c)(obj^{(6)},c)(obj^{(5)},u),0.5\}$

Since the certain part of each fuzzy incomplete equivalence class for attribute DP is not included in $X_H$, thus:

$DP_*(X_H) = \phi$

Similarly, the fuzzy incomplete lower approximations of single attributes for $X_N$ and $X_L$ are found as follows:

$SP_*(X_N) = \phi$

$DP_*(X_N) = \phi$

$SP_*(X_L) = \{(obj^{(4)},c)(obj^{(7)},c),0.5\}$

$DP_*(X_L) = \{(obj^{(4)},c)(obj^{(7)},u),1\}$

Step 6: Each uncertain object in the above fuzzy incomplete lower approximations is checked for change to certain objects. For example in $SP_*(X_H)$, since $Obj^{(5)}$ exist in only one fuzzy incomplete equivalence class of SP = H, their values can then be estimated from the certain objects in the same equivalence class. Since only one certain object $Obj^{(2)}$ and $Obj^{(6)}$ exists in the fuzzy incomplete equivalence class of SP = H, the estimated value of $Obj^{(5)}$ is then (155*0.75+150*0.5)/1.25 (=153), where 153 is the quantitative value of $Obj^{(2)}$ and $Obj^{(6)}$ for attribute SP and 0.75, 0.5 are there fuzzy membership values for the region of SP = H. The estimated values of $Obj^{(5)}$ is then transformed as the fuzzy set (0.2/N+0.65/H). $(Obj^{(5)},u)$ is then changed as $(Obj^{(5)},c)$.

The modified $SP_*(X_H)$ and The fuzzy incomplete elementary set of attribute SP are then:

$SP_*(X_H) = \{(obj^{(2)},c)(obj^{(6)},c)(obj^{(5)},c),0.5\}$

$U/\{SP\} = \{\{(obj^{(1)},c)(obj^{(2)},c)(obj^{(3)},c)$
$(obj^{(6)},c)(obj^{(7)},c)(obj^{(5)},c),0.1\}$
$\{(obj^{(2)},c)(obj^{(6)},c)(obj^{(5)},c),0.5\}$
$\{(obj^{(4)},c)(obj^{(7)},c),0.5\}\}$

Similarly, Since the uncertain object $Obj^{(7)}$ in $DP_*(X_L)$ exist in only the fuzzy incomplete equivalence class of DP = L, the estimated value of $Obj^{(7)}$ is then (68*1)/1 (=68)), where 68 is the quantitative value of $Obj^{(4)}$ for attribute DP and 1 is its fuzzy membership value of DP = L. The estimated value of $Obj^{(7)}$ is then transformed as the fuzzy set (1/L) for attribute DP. Also, $(Obj^{(7)},u)$ is then changed as $(Obj^{(7)},c)$. The modified $DP_*(X_H)$ and The fuzzy incomplete elementary set of attribute DP are then:

$DP_*(X_L) = \{(obj^{(4)},c)(obj^{(7)},c),1\}$

$U/\{DP\} = \{\{(obj^{(1)},c)(obj^{(2)},c)$
$(obj^{(3)},c)(obj^{(5)},c),0.16\}$
$\{(obj^{(3)},c)(obj^{(5)},c)(obj^{(6)},c),0.4\}$
$\{(obj^{(4)},c)(obj^{(7)},c),0.5\}\}$

Step 7: q = q+1, and Steps 5–7 are repeated. Until the fuzzy incomplete elementary set of attributes {SP, DP} is found as follows:

$U/\{SP,DP\} = \{\{(obj^{(1)},c)(obj^{(2)},c)$
$(obj^{(3)},c)(obj^{(5)},c),0.1\}$
$\{(obj^{(3)},c)(obj^{(5)},c)(obj^{(6)},c),0.2\}$
$\{(obj^{(2)},c)(obj^{(5)},c),0.16\}$
$\{(obj^{(4)},c)(obj^{(7)},c),0.5\}\{(obj^{(7)},c),0.1\}$
$\{(obj^{(5)},c)(obj^{(6)},c),0.5\}\}$

Because, It is requirement for next steps.

Step 8: Since all objects in the fuzzy incomplete lower approximations have become certain, go to the step12 is executed.

Step 12 : Initialize l=l+1, where l is used to count the number of classes then begin finding β-certain and β-possible rules processed.

with the relative degree β = 0.2 of misclassification can be calculated as follows:

for $X_N$ :

$c(SP_N(x), X_N) = 1 - \dfrac{0.9+0.85}{0.9+0.1+0.2+0.3+0.1+0.85} = 0.28$

$c(SP_H(x), X_N) = 1 - 0 = 1$

$c(SP_L(x), X_N) = 1 - 0 = 1$

$c(DP_N(x), X_N) = 1 - \dfrac{0.9+0.3}{0.4+0.9+0.3+0.16} = 0.31$

$c(DP_H(x), X_N) = 1 - \dfrac{0.4}{0.6+1+0.4} = 0.8$

$c(DP_L(x), X_N) = 1 - 0 = 1$

$c(SP, DP_{NN}(x), X_N) = 1 - \dfrac{0.9+0.3}{0.9+0.1+0.16+0.3} = 0.18$

$c(SP, DP_{HN}(x), X_N) = 1 - 0 = 1$

$c(SP, DP_{NH}(x), X_N) = 1 - \dfrac{0.4}{0.2+0.3+0.4} = 0.55$

$c(SP, DP_{LL}(x), X_N) = 1 - 0 = 1$

$c(SP, DP_{NL}(x), X_N) = 1 - 0 = 1$

$c(SP, DP_{HH}(x), X_N) = 1 - 0 = 1$

Step 13 : Assume β =0.2. The fuzzy β-lower and β-upper approximation for class $X_N=(\{Obj^{(1)},Obj^{(3)}\})$ is first calculated. Since only $\{Obj^{(1)},Obj^{(3)}\}$ is included in $X_N$, thus:

$SP,DP_{*0.2}(X_N) = \{\{(Obj^{(1)},Obj^{(2)},Obj^{(3)},Obj^{(5)}),0.1\}$

Similarly, the modified fuzzy β-upper approximations of single attributes for class $X_N$ is then calculated. Since only $\{Obj^{(1)},Obj^{(3)}\}$ is included in $X_N$, thus:

$SP^*_{0.2}(X_N) = \{(Obj^{(1)},Obj^{(2)},Obj^{(3)},Obj^{(5)},Obj^{(6)},Obj^{(7)}),0.1\}$

$DP^*_{0.2}(X_N) = \{\{(Obj^{(1)}, Obj^{(2)}, Obj^{(3)}, Obj^{(5)}), 0.16\} \{(Obj^{(3)}, Obj^{(5)}, Obj^{(6)}), 0.4\}\}$

$SP, DP^*_{0.2}(X_N) = \{\{(Obj^{(3)}, Obj^{(5)}, Obj^{(6)}), 0.2\}$

Step 15 Each partition in the fuzzy β-lower approximation is used to derived a β-certain rule with plausibility measure and future effectiveness measure. From $SP_{*\ 0.2}(X_N)$, the following rule is derived:

1. If Systolic Pressure = Normal and Diastolic Pressure = Normal Then Blood Pressure = Normal, with plausibility=0.82 and future effectiveness = 0.1.

Similarly, not exist rules are derived for $DP_{*0.2}(X_N)$ and $SP, DP_{*0.2}(X_N)$.

Step 16: Since the condition part of this rule is not more specific than any other rule in the β-certain rule sets, it is then kept in the fuzzy β-certain rule set.

Step 17 Each partition in the fuzzy β-upper approximation is used to derived a β-possible rule with plausibility measure and future effectiveness measure. From $SP^*_{0.2}(X_N)$, the following rule is derived:

2. If Systolic Pressure = Normal Then Blood Pressure = Normal, with plausibility=0.72 and future effectiveness = 0.1.

Similarly, the following rules are derived for $DP^*_{0.2}(X_N)$ and $SP, DP^*_{0.2}(X_N)$:

3. If Diastolic Pressure = Normal Then Blood Pressure = Normal, with plausibility =0.6 and future effectiveness = 0.16.
4. If Diastolic Pressure = High Then Blood Pressure =Normal, with plausibility =0.2 and future effectiveness = 0.4.
5. If Systolic Pressure = Normal and Diastolic Pressure = High Then Blood Pressure = Normal, with plausibility =0.45 and future effectiveness = 0.2.

Step 18: Since the condition part of this rule is not more specific than any other rule in the β-possible rule sets, it is then kept in the fuzzy β-possible rule set.

Step 19: Steps 13–19 are repeated to find rules for classes $X_L$ and $X_H$ until $l \leq c$.

Similarly, perform for $X_L$ and $X_H$. thus,

6. If Systolic Pressure = Low Then Blood Pressure = Low, with plausibility =1 and future effectiveness = 0.5.
7. If Diastolic Pressure = Low Then Blood Pressure = Low, with plausibility = 1 and future effectiveness = 1;
8. If Systolic Pressure =Low and Diastolic Pressure = Low Then Blood Pressure = Low, with plausibility =1 and future effectiveness = 0.5.
9. If Systolic Pressure = Normal and Diastolic Pressure = Low Then Blood Pressure = Low, with plausibility=1 and future effectiveness = 0.1.
10. If Systolic Pressure = High Then Blood Pressure = High, with plausibility =1 and future effectiveness = 0.5.
11. If Diastolic Pressure = High Then Blood Pressure = High, with plausibility = 0.8 and future effectiveness = 0.4;
12. If Systolic Pressure = High and Diastolic Pressure = Normal Then Blood Pressure = High, with plausibility =1 and future effectiveness = 0.16.
13. If Systolic Pressure = High and Diastolic Pressure = High Then Blood Pressure = High, with plausibility=1 and future effectiveness = 0.5.
14. If Diastolic Pressure = Normal Then Blood Pressure = High, with plausibility = 0.32 and future effectiveness = 0.16;
15. If Systolic Pressure = Normal and Diastolic Pressure = High Then Blood Pressure = High, with plausibility =0.56 and future effectiveness = 0.2.
16. If Systolic Pressure = Normal Then Blood Pressure = High, with plausibility=0.25 and future effectiveness = 0.1.

Since the condition part of the eightieth rule is more specific and both its plausibility (1) and effectiveness (0.5) is equal to those of the sixth rule, this rule is thus removed from the β-certain rule set. Similarly, rules 9,12 and 13 are also removed by fuzzy rules 7 and 10.

Step 20 : all the fuzzy β-certain and β-possible rules are shown below:

Fuzzy β-certain rules :

1. If Systolic Pressure = Normal and Diastolic Pressure = Normal Then Blood Pressure = Normal, with plausibility=0.82 and future effectiveness = 0.1.
2. If Systolic Pressure = Low Then Blood Pressure = Low, with plausibility =1 and future effectiveness = 0.5.
3. If Diastolic Pressure = Low Then Blood Pressure = Low, with plausibility = 1 and future effectiveness = 1;
4. If Systolic Pressure = High Then Blood Pressure = High, with plausibility =1 and future effectiveness = 0.5.
5. If Diastolic Pressure = High Then Blood Pressure = High, with plausibility = 0.8 and future effectiveness = 0.4.

Fuzzy β-possible rules :

1. If Systolic Pressure = Normal Then Blood Pressure = Normal, with plausibility=0.72 and future effectiveness = 0.1.
2. If Diastolic Pressure = Normal Then Blood Pressure = Normal, with plausibility =0.6 and future effectiveness = 0.16.
3. If Diastolic Pressure = High Then Blood Pressure = Normal, with plausibility =0.2 and future effectiveness = 0.4.

4. If Systolic Pressure = Normal and Diastolic Pressure = High Then Blood Pressure = Normal , with plausibility =0.45 and future effectiveness = 0.2.

5. If Diastolic Pressure = Normal Then Blood Pressure = High, with plausibility = 0.32 and future effectiveness = 0.16;

6. If Systolic Pressure = Normal and Diastolic Pressure = High Then Blood Pressure = High, with plausibility =0.56 and future effectiveness = 0.2 .

7. If Systolic Pressure = Normal Then Blood Pressure = High , with plausibility=0.25 and future effectiveness = 0.1.

## VII. Discussion and conclusion

In this paper, we have proposed a novel data mining algorithm, which can process incomplete quantitative data with a predefined tolerance degree of uncertainty and misclassification. The interaction between data and approximations helps derive $\beta$-certain and $\beta$-possible rules from fuzzy incomplete data sets and estimate appropriate unknown values. The fuzzy $\beta$-certain rules with misclassification degrees smaller than $\beta$ and the fuzzy $\beta$-possible rules with misclassification degrees smaller than $1-\beta$ are derived. Noisy training examples (as outliers) may then be omitted. The rules thus mined exhibit fuzzy quantitative regularity in databases and can be used to provide some suggestions to appropriate supervisors. The proposed algorithm can also solve conventional incomplete quantitative data problems by using degraded membership functions. The selection of $\beta$ values is remarked here. When $\beta = 0$, the proposed approach will be reduced to the one for the original rough-set model with incomplete quantitative data. A larger $\beta$ value represents a larger tolerance of uncertainty and misclassification, but with a smaller gap between lower approximations and upper approximations. The selection of an appropriate $\beta$ value then depends on given training instances.

## Acknowledgement

This research was supported by the affiliation must be Islamic Azad University, Arak Branch of Iran.